\documentclass[twocolumn,superscriptaddress,showpacs]{revtex4-1}

\usepackage{amssymb}
\usepackage{amsmath}
\usepackage{graphicx}

\def\g{{\bf{g}}}

\def\g0{{\gamma_0}}

\def\Im{{\mbox{Im}}}

\def\figwidth{0.9\columnwidth}

\begin{document}

\title{Time dependent local potential in a Tomonaga-Luttinger liquid}
\author{Naushad Ahmad Kamar and Thierry Giamarchi}
\affiliation{DQMP, University of Geneva, 24 Quai Ernest-Ansermet, 1211 Geneva, Switzerland}

\begin{abstract}
We study the energy deposition in a  one dimensional interacting quantum system with a point like potential modulated in amplitude. The point like potential at position $x=0$ has a constant part and a small oscillation in time with a frequency $\omega$. We use bosonization, renormalization group and linear response theory to calculate the corresponding energy deposition. It exhibits a power law behavior as a function of the frequency that reflects the Tomonaga-Luttinger liquid (TLL) nature of the system. Depending on the interactions in the system, characterized by the TLL parameter $K$ of the system, a crossover between week and strong coupling for the backscattering due to the potential is possible. We compute the frequency scale $\omega_\ast$, at which such crossover exists. We find that the energy deposition due to the backscattering shows different exponent for $K>1$ and $K<1$. We discuss possible experimental consequences, in the context of cold atomic gases, of our theoretical results.
\end{abstract}
\pacs{03.75.Kk, 05.30.Jp, 03.75.Lm, 73.43.Nq}
\maketitle

\section{Introduction}

Cold atomic systems have proven in the recent year to be a remarkable playground to study the effect of strong correlations in quantum systems\cite{bloch_2008_ultracold_system,esslinger_annrev_2010}.
In particular they have allowed the realization of one dimensional quantum systems both for bosons and for fermions. In one dimension interactions among the particles lead to a very rich set of properties, very different from their higher dimensional counterparts. The general physics goes by the description known as Tomonaga-Luttinger liquid (TLL)\cite{Haldane_TLL_Theory,GOGOLIN_Bosonization,giamarchi_book_1d,cazalilla_review_bosons}.

In a TLL, the interactions lead to a critical state in which correlations decrease as power laws. This make the system extremely sensitive to external perturbations, such as a periodic
potential \cite{georges_leshouches_cold_lectures}, quasi-periodic and disordered potentials \cite{TG_disordered_Luttinger_Liquid,TG_Quasiperiodic_Bose_Hubbard_model} and even the presence of a single impurity \cite{kane_luttinger_impurity,furusaki_luttinger_impurity}.
In the later case depending on the interactions in the system even a single impurity can potentially completely block the system.
Such effects have been investigated in the condensed matter context both
in the context of carbon nanotubes \cite{Yao_impurity_Luttinger_Liquid} and also of edge states in the quantum Hall effect \cite{Fendley_Transport_Through_Quantum_Point_Contact_in_Hall_Effect}. In such context the main probe is connected to the transport through the impurity site, and the control parameters are either the temperature or the voltage.

In the present paper we propose to investigate the possibilities that are offered by the remarkable tunability and control of cold atom realization to study this class of effects.
In particular we consider the question of a local potential, present in a one dimensional system, that would be periodically modulated in time around an average value.
Such a technique of periodically modulating potentials has been exploited with success to investigate several aspects of the Mott transition both in bosonic and fermionic systems \cite{stoferle_tonks_optical,schori_absorption_optical}, for disordered systems
\cite{derrico_cold_boseglass} or for superconducting systems \cite{Bloch_Higgs_mode_cold_atoms} .
A theoretical analysis shows that measuring quantities such as the deposited energy \cite{iucci_absorption,tokuno_sieving_conductivity} or the raise of the number of doubly occupied sites \cite{Kollath_Phase_Modulation,Tokuno_Doublon_production_rate_in_modulated_optical_lattices}, in response to a periodic global modulation provides a spectroscopic probe akin, but not alike to the optical conductivity. We show here that the analysis of the deposited energy in the system as a function of the frequency dependence of the \emph{local} modulation gives direct information
on the physics of the static problem with the impurity \cite{kane_luttinger_impurity,furusaki_luttinger_impurity}. The frequency dependence of the energy absorbed is a power law reflecting
the TLL nature of the system.

The paper is structured as follow: in Sec.~\ref{sec:model}, we introduce the model we use, within the framework of TLL to study the
energy deposition by using linear response theory (LRT). Depending on the interaction two cases in which the impurity potential would be relevant (resp. irrelevant) in the static
case must be distinguished. Sec.~\ref{sec:weakcoupling} discusses the deposited energy for the case in which the static impurity potential is irrelevant. This typically corresponds to $K>1$ for the TLL. The opposite case of $K<1$ is examined in Sec.~\ref{sec:strongcoupling}. The results and their potential application to experimental systems in cold atomic gases are discussed in Sec.~\ref{sec:discussion}. Conclusion is given in Sec.~\ref{sec:conclusion} and some technical details are described in the appendices.

\section{Model and Method} \label{sec:model}

\subsection{Model}

\begin{figure}
\begin{center}
 \includegraphics[width=\figwidth]{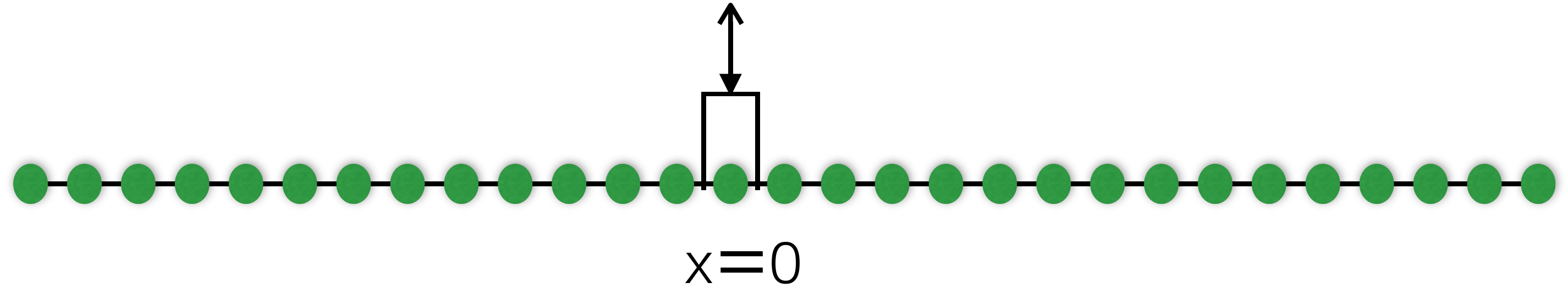}
\caption{(color online) Sketch of a one dimensional system with a time dependent external local potential $V(x=0,t)=V_0+\delta V \cos(\omega t)$. The rectangular box represents a static potential $V_0$. The double head arrow represents the time dependent part of the potential $\delta V \cos(\omega t)$.}
\label{fig:fig1}
\end{center}
\end{figure}

We consider a one dimensional system, schematically shown in Fig.~\ref{fig:fig1}.
The system is made of  interacting bosons or fermions described by a Hamiltonian $H_0$ which depends on
the precise system under consideration (e.g. continuum or system on a lattice) and will be made more precise below.
Most of this paper is concerned with a single component system, of bosons or fermions, but we also discuss briefly the case of a two component fermionic system.
This is particularly relevant in connection with recent realizations of fermionic microscope \cite{Zwierlein_Fermion_Microscope,Kuhr_Fermion_Microscope} that allow the label of control relevant to carry the time of modulation and measurements corresponding to the present study. We consider separately the case of weak and strong potentials for a single component case and then address the two component weak potential case.

We add to this system a time dependent potential $V(x,t)$. The total Hamiltonian we consider is thus:
\begin{equation}\label{eq:eq0}
 \begin{split}
  H &= H_0 + H_{\text{imp}}\\
  H_{\text{imp}} &= \int dx V(x,t)\rho(x)
 \end{split}
\end{equation}
where $\rho(x)$ represents the density of particles at position $x$. We compute here the change in energy caused by this \emph{local} modulation.
This corresponds to the local version of the shaking technique, in which a full optical lattice is modulated \cite{iucci_absorption,Kollath_Phase_Modulation,tokuno_sieving_conductivity}.

The  modulation of the amplitude of the local potential is given by (see Fig.~\ref{fig:fig1})
\begin{equation} \label{eq:vertpot}
 V(x,t) = [V_0 + \delta V \cos(\omega t)] f_\lambda(x)
\end{equation}
where $f_\lambda(x)$ is a short range function which is a regularized $\delta$ function. For simplicity we chose
\begin{equation}
 f_\lambda(x) = \frac{1}{2 \lambda \cosh^2(x/\lambda)}
\end{equation}
the $\delta$ function being reproduced in the limit $\lambda\to 0$.

Thus  for (\ref{eq:vertpot})  the impurity potential can be split in a purely static part
corresponding to the one of a single impurity and a purely dynamical part, of mean zero. We can thus quite generally write
\begin{equation} \label{eq:general}
 H = H_S + \cos(\omega t) O
\end{equation}
where $H_S$ is the purely static impurity potential, and $O$ the operator corresponding to the modulation given by (\ref{eq:vertpot}).

If the impurity potential $V_0$  is weak compared to the kinetic and interaction energy of the system, then
one has
\begin{equation}
 H_S = H_0 + V_0 \int dx f_\lambda(x) \rho(x)
\end{equation}
We consider that $f_\lambda(x)$ varies fast enough compared
to the typical scales of the problem and thus can be assimilated to a $\delta(x)$ function as far as the density variation of
the system are concerned.

Since the \emph{perturbation} is small we use linear response theory. The energy deposition rate is given by(see appendix \ref{sec:linrep})
\begin{equation}\label{eq:eq15}
 E_R = -\frac{1}{2} \omega \Im\chi^R(\omega)
\end{equation}
where
\begin{equation}
 \chi^R(\omega) = -i \int_0^\infty dt e^{i(\omega+i\delta) t} \langle [O(t),O(0)] \rangle
\end{equation}
is the retarded correlation function of the operator $O$ linked to the perturbation (see Appendix~\ref{sec:linrep}).
$\langle \cdots \rangle$ denotes the quantum and/or thermal average with the Hamiltonian $H_S$ in (\ref{eq:general}).

\subsection{Weak potential, bosonization}\label{sec:weak_potential}

In order to treat the Hamiltonian (\ref{eq:general}) we use the bosonization technique. Indeed for one dimensional quantum systems and regardless of the statistics of the particles, the
low energy excitations of the system can be represented by the Hamiltonian \cite{giamarchi_book_1d}:
\begin{equation}\label{eq:eq1}
H_0 = \frac{1}{2\pi}\int dx\Big[u K(\partial_x \theta(x))^2+\frac{u}{K} (\partial_x \phi(x))^2 \Big]
\end{equation}
The fields $\phi$ and $\theta$ are canonically conjugate
\begin{equation} \label{eq:canonical}
  [\phi(x),\frac1\pi\nabla\theta(x')] = i\delta(x-x')
\end{equation}
The field $\phi$ is related to the density of particles via
\begin{equation} \label{eq:density}
 \rho(x) = \rho_0 -\frac1\pi\nabla\phi(x) + \rho_0 \sum_{p\neq 0} e^{i 2 p(\pi \rho_0 x - \phi(x))}
\end{equation}
where $\rho_0$ is the average density of particles. Note that for fermions $2\pi\rho_0 = 2 k_F$ where $k_F$ is the Fermi wavevector.
The above formula are the basis for the so-called bosonized representation, that relates the original Hamiltonian in terms of collective variable of
density ($\phi$) and current ($\theta$). The transformation is by now standard and we refer the reader to the literature for more details \cite{giamarchi_book_1d}.

In (\ref{eq:eq1}) $u$ and $K$ are known as Tomonaga-Luttinger liquid parameters and they depend in general on the properties of the system and in particular on the interaction
between the particles. $u$ is the velocity of sound excitations, while $K$ is a dimensionless parameter controlling the decay of the correlation functions.
For fermions, $K=1$ for noninteracting particles while $K>1$ (resp. $K<1$) for attraction (resp. repulsion) between the particles.
For bosons $K=\infty$ for noninteracting particles, and becomes smaller as the repulsion increases. For contact interaction only $K>1$, with $K=1$ being reached
for infinite repulsion between the bosons (the so-called Tonks-Girardeau limit). Longer range interactions allow in principle to reach any value of $K$.
For special models such as the Lieb-Lininger model or the Bose-Hubbard one for bosons or the Hubbard model for fermions precise relations between the microscopic
parameters and the TLL parameters are known \cite{giamarchi_book_1d,cazalilla_review_bosons}. In general if the microscopic parameters are known the TLL parameters can be computed with
an arbitrary degree of accuracy. We thus use the Hamiltonian (\ref{eq:eq1}) and (\ref{eq:general}) with the representation (\ref{eq:density}) as our starting model
in this paper.

The Hamiltonian $H_S$ thus describes a TLL with a single impurity, for which physics is well known \cite{kane_luttinger_impurity}.
We briefly recall the main elements of this Hamiltonian since it will be needed to compute the effects of the dynamical perturbation.

For a perturbation weak compared to the kinetic and interaction energy, using (\ref{eq:density}) the impurity part becomes
\begin{equation} \label{eq:tllimp}
\begin{split}
 H^0_{\text{imp}} &= V_0 \int dx f_\lambda(x) \rho(x) \\
     &= V_0 \int dx f_\lambda(x) \left[ -\frac1\pi \nabla\phi(x) + 2 \rho_0 \cos(2\phi(x)) \right]
\end{split}
\end{equation}
where we have kept only the lowest ($p=\pm 1$) and the most relevant harmonics in density (\ref{eq:density}).

\begin{figure}
\begin{center}
 \includegraphics[width=\figwidth]{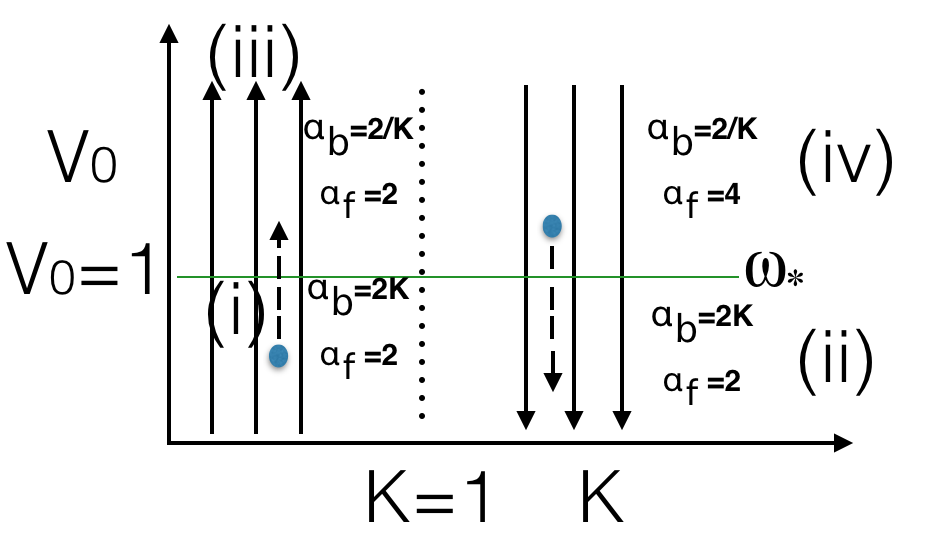}
\caption{(color online) Various regimes for the energy absorption as a function of the TLL parameter $K$ and the strength of the backscattering potential $V_0$. The solid line arrows indicate
the renormalization flow of the static backscattering potential $V_0$ \cite{kane_luttinger_impurity,furusaki_luttinger_impurity}. For $K<1$, $V_0$ flows to the strong coupling limit, and for $K>1$, $V_0$ flows to zero. The green line represents the crossover scale $\omega_\ast$ at which the renormalized potential is of order one.
$\alpha_b$ and $\alpha_f$ represent the energy deposition exponents corresponding to the backward and forward scattering respectively (see text). The solid blue circle represents the initial point for the local potential. For $K<1$, even if the initial point corresponds to a small  local potential, it renormalizes to strong coupling when lowering the frequency $\omega$ but for $K>1$, even if the initial point is a large local potential it renormalizes to  weak potential via the lowering of $\omega$. }
\label{fig:fig2}
\end{center}
\end{figure}

The first term can be absorbed by a redefinition of the field $\phi$
\begin{equation} \label{eq:redefphi}
 \tilde\phi(x) = \phi(x) -\frac{KV_0}{u} \int_0^x dy f_\lambda(y)
\end{equation}
The $\theta$ field is not affected by this transformation which preserves the canonical commutation relations (\ref{eq:canonical}).
In the limit where $\lambda\to 0$ the cosine term is not affected by this transformation since it contains
only the value $\phi(x=0)$. The static Hamiltonian is thus
\begin{equation} \label{eq:tllstat}
 H_S = H_0[\tilde\phi] + 2 V_0 \rho_0 \cos(2\tilde\phi(0))
\end{equation}
The ground state depends crucially on the value of $K$ \cite{kane_luttinger_impurity,furusaki_luttinger_impurity}. A renormalization
group procedure based on changing the cutoff in the problem (e.g. a high energy cutoff $\Lambda$) gives the flow of the parameters
\begin{equation}\label{eq:eq6}
\begin{split}
\frac{dK}{dl}&=0\\
\frac{dV_0}{dl}&=(1-K)V_0
\end{split}
\end{equation}
where $l$ is the scale of the renormalization for an effective high energy cutoff $\Lambda_l = \Lambda e^{-l}$, $\Lambda$ is the maximum high energy cutoff, of the order of the bandwidth.
Thus for $K>1$, $V_0$ flows to $0$ and the cosine term is irrelevant. On the other hand for $K<1$ the cosine is relevant and $V_0$ flows to the the strong coupling limit.
A schematic view of the RG flow is shown in Fig.~\ref{fig:fig2}.

The redefinition of the field $\phi$ (\ref{eq:redefphi}) can also be done in the dynamical terms. Up to terms that are simply oscillating and thus do not contribute to the increase of the
average energy (\ref{eq:eq15}) one obtains in the limit $\lambda \to 0$
\begin{equation} \label{eq:moduv}
 O_V = \delta V \left[-\frac1\pi \nabla\tilde\phi(x=0) + 2 \rho_0 \cos(2\tilde\phi(x=0))\right]
\end{equation}

\subsection{Two component fermionic system}

In this section, we consider a two component fermionic system, given by the Hubbard model. We use similar techniques as for the single component case.
We restrict ourselves for simplicity to the weak coupling case.

The static part of Hamiltonian in bosonized language is given by \cite{giamarchi_book_1d}
\begin{multline} \label{eq:fermionic1}
 H_S = H_{0}[\phi_\rho] +H_{0}[\phi_\sigma] \\ + 2 V_0 \rho_0 \cos(2\phi_{\uparrow}(0))+ 2 V_0 \rho_0 \cos(2\phi_{\downarrow}(0))
\end{multline}
\begin{equation} \label{eq:fermionic2}
 H_S = H_{0}[\phi_\rho] +H_{0}[\phi_\sigma] + 4 V_0 \rho_0 \cos(\sqrt 2\phi_{\rho}(0))\cos(\sqrt 2\phi_{\sigma}(0))
\end{equation}
Where $\phi_\uparrow$ and $\phi_\downarrow$ are fields for spin up and spin down fermions, $\phi_\rho=\frac{\phi_\uparrow+\phi_\downarrow}{\sqrt 2}$, $\phi_\sigma=\frac{\phi_\uparrow-\phi_\downarrow}{\sqrt 2}$ and
\begin{equation}\label{eq:fermionic3}
H_{0}[\phi_\rho] = \frac{1}{2\pi}\int dx\Big[u_\rho K_\rho(\partial_x \theta_\rho(x))^2+\frac{u_\rho}{K_\rho} (\partial_x \phi_\rho(x))^2 \Big]
\end{equation}
\begin{equation}\label{eq:fermionic4}
H_{0}[\phi_\sigma]= \frac{1}{2\pi}\int dx\Big[u_\sigma K_\sigma(\partial_x \theta_\sigma(x))^2+\frac{u_\sigma}{K_\sigma}(\partial_x \phi_\sigma(x))^2 \Big]
\end{equation}
$u_\rho, u_\sigma, K_\rho, K_\sigma$ are the Luttinger parameters. For a repulsive interaction $K_\rho<1$ and for a spin rotation symmetric Hamiltonian $K_\sigma$ renormalizes to
$K_\sigma^*=1$.

The renormalization flow of $V_0$ is given by \cite{kane_luttinger_impurity,furusaki_luttinger_impurity}
\begin{equation}\label{eq:fermionic5}
\frac{dV_0}{dl}=[1-(K_\sigma+K_\rho)/2]V_0
\end{equation}
$V_0$ is irrelevant for $K_\sigma+K_\rho>2$, and relevant for $K_\sigma+K_\rho<2$.
The time dependent part of the Hamiltonian for a weak potential is given by
\begin{multline} \label{eq:fermionic6}
 O_V = \delta V \left[-\frac {\sqrt 2}{\pi} \nabla\phi_\rho(x=0)\right. \\
  \left. +4 V_0 \rho_0 \cos(\sqrt 2\phi_{\rho}(0))\cos(\sqrt 2\phi_{\sigma}(0))\right]
\end{multline}

\subsection{Strong coupling}\label{strong_coupling}

The description of the previous two subsections is well adapted if the potential $V(x,t)$ (and the modulation itself) are weak compared to other parameters in the problem,
such as e.g. the kinetic energy. If the impurity potential $V$ is the largest scale in the problem, then it is more useful to write an effective description using an expansion in powers of $1/V$.
We examine such a case for a single component fermionic system.

In order to do so, we put the system on a lattice and assume that the potential only exists on the site $x=0$. We also assume that the interactions are onsite and nearest neighbour.
We divide the full Hamiltonian in three parts. The first (resp. second) part contains all sites left (resp. right) of the origin. The third part is the impurity site.
The tunnelling and nearest neighbour interaction between sites (kinetic energy) connects the three parts:
\begin{equation}\label {eq:barrier1}
 H = H_1 + H_2 - J [\psi_{-1}^\dagger \psi_0+ \psi_{1}^\dagger \psi_0+ h.c] + V \psi_0^\dagger \psi_0 + V_n n_0(n_{-1}+n_1)
\end{equation}
For a large $V$ we derive the effective Hamiltonian in powers of $1/V$ (see Appendix~\ref{sec:effective hamiltonian}). The effective Hamiltonian is
\begin{equation}\label{eq:barrier2}
\begin{split}
 H &= H_1 + H_2 - \frac{ J^2}{V}[(\psi^\dagger_{1} \psi_{-1} + \\& \text{h.c.})+\psi^\dagger_{1} \psi_{1} +\psi^\dagger_{-1} \psi_{-1}]
 \end{split}
\end{equation}
We thus see that the two half parts are coupled by an effective tunneling term
\begin{equation}
 J^{\rm eff} = \frac{J^2}{V}
\end{equation}
and that there is an attractive potential of the same order acting on the last site of each half part.

For the dynamical case we simply assume that we can replace the static $V$ by the dynamical one
$V=V_0+\delta V \cos(\omega t)$, since $V_0 \gg 0$, $1/V = 1/V_0-\delta V/V_0^2\cos(\omega t)$. The static part of the Hamiltonian is given by
\begin{equation} \label{eq:barrier}
\begin{split}
 H_S &= H_1 + H_2 - J_0 [(\psi^\dagger_{-1}(x=-a) \psi_{1} + \text{h.c.})+\\& \psi^\dagger_{1}\psi_{1} +\psi^\dagger_{-1}\psi_{-1} ]
 \end{split}
\end{equation}
Where $1$ and $2$ stand for right and left part of one dimensional chain, $J_0= J^2/V_0$ and $a$ is lattice constant.

In this case the amplitude modulation of the potential produces a small modulation of the tunnelling amplitude and  local potential.
The operator $O$ is
\begin{equation} \label{eq:modbar}
\begin{split}
 O &= \delta t_1 (\psi^\dagger_{1}(x=-a) \psi_{2}(x=a) + \text{h.c.})+\\& \delta t_2 (\psi^\dagger_{2}(x=a) \psi_{2}(x=a) +\psi^\dagger_{1}(x=-a) \psi_{1}(x=-a)  )
 \end{split}
\end{equation}
where $\delta t_2=-\frac{ \delta V J^2}{ V_0^2}$, and $\delta t_1=-\frac{ \delta V J^2}{ V_0^2}$.
Since $ J^2/V_0$ is small compared to the kinetic energy we can bosonize (\ref{eq:barrier2}). The expressions are derived in Appendix~\ref{sec:effective hamiltonian}, and the final static Hamiltonian is given by
\begin{equation}\label {eq:largeV_hamiltonian}
\begin{split}
H_s &= \frac{u}{2\pi}\int_{-\infty} ^{\infty} dx\Big[ (\partial_x \tilde{\tilde{\theta}}(x))^2+ (\partial_x \tilde{\tilde{\phi}}(x))^2\Big] +\\& \frac {2 a \sqrt{K}J^2}{\pi V_0} \partial_{x}^2 \tilde{\tilde{\theta}}(0) -\frac {2J^2\rho_0}{ V_0}\cos\Big(\frac{\sqrt K}{2}( \tilde{\tilde{\theta}}(a)-\tilde{\tilde{\theta}}(-a))\Big)\\& \cos\Big(\frac{\sqrt K}{2}( \tilde{\tilde{\phi}}(a)-\tilde{\tilde{\phi}}(-a))\Big)-  \frac{2 J^2\rho_0}{V_0}\cos\Big(\frac{ 2}{\sqrt K}( \tilde{\tilde{\phi}}(0))\Big)
\end{split}
\end{equation}

\section{Calculation of the energy absorption}

We now compute the energy absorption for the amplitude modulation.
Let us consider first the case for which the bare potential is weak as described in Sec.~\ref{sec:weak_potential}.

\subsection{Weak bare potential}

If one starts with a weak initial potential both the Hamiltonian and the perturbation are described by (\ref{eq:eq1}) and (\ref{eq:moduv}) for the single component case and by (\ref{eq:fermionic1}) and (\ref{eq:fermionic6}) for the two component one.
In the static part of the Hamiltonian the backscattering part of the potential renormalizes as described by (\ref{eq:eq6}) and (\ref{eq:fermionic5}).
Depending on the value of $K$ (or $K_\rho$ for the two component case) the potential can either renormalize to zero or flow to strong coupling, leading to a
crossover to another regime. If $K > 1$ (resp. $K_\rho > 1$ for the two component) the backscattering term renormalizes to zero, and the energy absorption can thus essentially
be computed with the quadratic part of the Hamiltonian (\ref{eq:eq1}), (\ref{eq:fermionic3}), (\ref{eq:fermionic4}).
On the contrary if $K < 1$ (resp. $K_\rho < 1$) the \emph{backscattering} term in (\ref{eq:tllstat}) and (\ref{eq:fermionic1}) renormalizes to large
values. In that case there is a crossover scale below which one cannot ignore the backscattering term in the static Hamiltonian.

We examine sequentially these two cases.

\subsection{Energy deposition in weak coupling limit} \label{sec:weakcoupling}

To compute the energy absorption rate for the single component case (\ref{eq:moduv}) we use (\ref{eq:eq15})
for a Hamiltonian $H = H_0 + \cos(\omega t) (a_1O_1+a_2O_2)$ where, using (\ref{eq:moduv})
\begin{equation}
 O_1 = \cos(2\phi(x=0))
 \end{equation}
\begin{equation}
 O_2 = \nabla\phi(x=0)
\end{equation}

This leads to
\begin{equation}\label{eq:energydepo}
\begin{split}
 E_R = -\frac{1}{2}a_1^{2} \omega  \mathrm{Im}[\chi_1^R(\omega)]-\frac{1}{2}a_2^{2} \omega  \mathrm{Im}[\chi_2^R(\omega)]
\end{split}
\end{equation}
where $\chi_1^R(t)=-i Y(t)\langle[O_1(t), O_1(0)]\rangle_{H_0} $,  $\chi_2^R(t)=-i Y(t)\langle[O_2(t), O_2(0)]\rangle_{H_0} $. $Y(t)$ is Heaviside step function.

The explicit calculation of the correlation is performed in Appendix~\ref{sec:correl} and gives
\begin{equation}\label{eq:eq18}
\begin{split}
E_R &=-\frac{1}{2}\Big(\frac{\delta V}{\pi}\Big)^{2} \omega  \mathrm{Im}[\chi_2^R(\omega)]-\frac{1}{2}(2\delta V\rho_0)^2\omega \mathrm{Im}[\chi_1^R(\omega)]\\
&=\frac{\delta V^2 K}{4 \pi u^2}\omega^2+\frac{1}{2}(2\delta V\rho_0)^2\sin(K\pi)\cos(K\pi)\\& \quad ({u/\alpha})^{-2K}\Gamma(1-2K)\omega^{2K}
\end{split}
\end{equation}
Where $\chi_1^R(\omega)$ and $\chi_2^R(\omega)$ are defined in (\ref{eq:eq8B}) and (\ref{eq:eq9B}) of the Appendix~\ref{sec:correl}, $\alpha$ is of order of lattice cutoff.\\

As we see the energy deposition rate $E_R$ shows a power law behavior as a function of the frequency of the potential modulation.
The forward scattering part of the potential leads to an exponent of two, while the backward scattering part has an exponent $2K$.
For $K > 1$ the energy deposition rate is dominated at small frequencies by the forward scattering part. Note that the prefactor of the energy absorption gives direct access to the
TLL parameters of the system.

For the two component case, similar calculations lead to
\begin{equation}\label{eq:fermionic8}
\begin{split}
E_R &=\frac{\delta V^2 K_\rho}{2 \pi u_\rho^2}\omega^2+\frac{1}{2}(4\delta V\rho_0)^2\sin((K_\rho+K_\sigma)\pi/2) \\& \cos((K_\rho+K_\sigma)\pi/2) (u_\rho/\alpha)^{-K_\rho}(u_\sigma/\alpha)^{-K_\sigma}\\& \Gamma(1-(K_\rho+K_\sigma))\omega^{K_\rho+K_\sigma}
\end{split}
\end{equation}
If $K_\rho > 1$ then the dominant contribution comes also from the forward scattering for small frequencies.

\subsection{Crossover from weak to strong coupling limit and strong to weak coupling limit}\label{sec:Crossover_from_weak_to_strong_coupling_limit_and_strong_to_weak_coupling_limit}

The behavior of the previous section remains valid only if one can compute the absorbed energy with the quadratic part of the static Hamiltonian only. This is valid for $K>1$ or $K_\rho>1$
since the bare potential of the backscattering term scales down, but becomes incorrect in the opposite limit of $K<1$ (or $K_\rho <1$) since in that case the
backscattering part scales up in the static Hamiltonian. The RG equations (\ref{eq:eq6}) and (\ref{eq:fermionic5}) thus define a scale at which the backscattering term renormalizes
to a value of order one, and thus one enters a different regime to compute the absorption. As a function of the frequency of the modulation this defines a crossover frequency $\omega_\ast$ below which the expressions (\ref{eq:eq18}) and (\ref{eq:fermionic8}) are not valid any more.

Let us start from a regime where $K<1$, and a weak bare impurity potential $V_0<1$. $V_0$ will flow to the strong coupling limit by lowering the frequency of the modulation.
We can compute the flow parameter $l_\ast$ at which $V_0(l_\ast)\simeq \alpha \ \omega_0$. Using (\ref{eq:eq6}) we obtain
\begin{equation} \label{eq:weak to strong3}
 V_0(l_{\ast})=V_0(l=0) \exp[(1-K) l_\ast] \simeq \alpha \ \omega_0
\end{equation}
\begin{equation} \label{eq:weak to strong4}
 \exp[ l_{\ast }]={\Big(\frac{\alpha \omega_0}{V_0(l=0)}\Big)}^{\frac{1}{1-K}}
\end{equation}
If $\omega_0$ is the maximum frequency cutoff of the order of the bandwidth
\begin{equation} \label{eq:weak to strong5}
 \omega_\ast=\omega_0 {\Big(\frac{\alpha \omega_0}{V_0(l=0)}\Big)}^{\frac{-1}{1-K}}
\end{equation}

In an opposite way, if we had started in the strong coupling regime for which the bare potential of the impurity is very strong as described in Sec.~\ref{strong_coupling} but have $K > 1$
the weak tunnelling $J_0$ between the two parts of the chain would flow according to \cite{kane_luttinger_impurity,giamarchi_book_1d}
\begin{equation} \label{eq: strong to weak1}
 \frac{dJ_0}{dl}=\left[(1-1/K)J_0
 \right]
\end{equation}
leading in a similar way to a crossover scale at which the tunnelling becomes of order one
\begin{equation} \label{eq: strong to weak5}
 \omega_\ast=\omega_0 {\Big(\frac{\omega_0}{J_0(l=0)}\Big)}^{\frac{-1}{1-1/K}}
\end{equation}

\subsection{Strong coupling renormalization of backscattering ($K<1$, $\omega<\omega_\ast$)} \label{sec:strongcoupling}

For frequencies of the modulation $\omega \ll \omega_\ast$ one enters (for $K<1$ or $K_\rho<1$) the strong coupling regime for which the backscattering term plays a central role
in the static Hamiltonian. Note that this regime is a priori different from the case for which one would have started from a strong bare potential. Indeed in that case both forward
and backward scattering would be affected. In the present case the forward scattering potential is not affected in the static part of the Hamiltonian.

To calculate the energy deposition, we use the dilute instanton approximation \cite{giamarchi_book_1d} as described in Appendix~\ref{ap:instanton}.
The energy deposition due to the amplitude modulation in the strong coupling regime is given by
\begin{equation}\label{eq:eq18a}
\begin{split}
E_R&=\frac{\delta V^2 K}{4 \pi u^2}\omega^2+\frac{1}{2}M^2(2\delta V\rho_0)^2 \cos(\pi/K)\sin(\pi/K)\\& \Gamma(1-2/K)(\delta\  \omega)^{2/K} e^{-8\sqrt{2\rho_0V_0(l_\ast)M}}
\end{split}
\end{equation}
Where $\delta$ is short time cutoff.
In the dilute instanton calculation, one assumes that the backward scattering potential is very large but near $\omega=\omega_\ast$, $V_0$ is not very large. The dilute instanton approximation gives the correct exponent of $E_R$ but does not provide the correct coefficient near $\omega_\ast$. $E_R$ in weak coupling and strong coupling limit should match at $\omega=\omega_\ast$.
Using the fact that for $\omega < \omega_\ast$ one has from (\ref{eq:eq18a}) $E_R = A (\omega \alpha/u)^{2/K}$ where $A$ is a coefficient to be determined, and for $\omega > \omega_\ast$ we have (\ref{eq:eq18}), the continuity equation reads 
\begin{equation} 
 A (\alpha/u)^{2/K}\omega_\ast^{2/K} = \sin(K\pi)\cos(K\pi) ({u/\alpha})^{-2K}\Gamma(1-2K)\omega_\ast^{2K}
\end{equation}
which determined the prefactor $A$ 
\begin{equation}\label{eq:eq18ab}
\begin{split}
A&=\sin(K\pi)\cos(K\pi)\\& \quad ({u/\alpha})^{2/K-2K}\Gamma(1-2K)\omega_\ast^{2K}/( \omega_\ast)^{2/K}\\
&=\sin(K\pi)\cos(K\pi) \quad ({u/\alpha})^{2/K-2K}\Gamma(1-2K)\\& \omega_0^{2(K-1/K)}(\alpha \omega_0/V_0)^{2(K+1)/K}
\end{split}
\end{equation}

At the frequency $\omega_\ast$ the ratio between the part of $E_R$ due to the backward scattering and the one due to the forward scattering is 
\begin{equation} 
r=\frac{4\pi u^2\rho_0^2 \alpha^2 \sin(2 K\pi)(\alpha\omega_0)^{2K}}{KV_0^2u^{2K}}
\end{equation} 
Note that for a weak initial potential this ratio can become very large. Thus for frequencies $\omega < \omega_\ast$ the backward scattering will continue to dominate down to a lower 
frequency $\omega_1$ below which the energy absorption rate will be dominated by the forward scattering. This frequency is given by 
\begin{equation}
 \omega=\omega_1=\left(\frac{\delta V^2 K}{4 \pi u^2 A}\right)^{1/(2(1/K-1))}
\end{equation}
Note that for $K<1$ and $\omega>\omega_\ast$, $\omega<1$ the energy deposition is still given by (\ref{eq:fermionic8}), and the dominant
contribution comes from the backward scattering.

\begin{figure}
\begin{center}
 \includegraphics  [width=\figwidth]{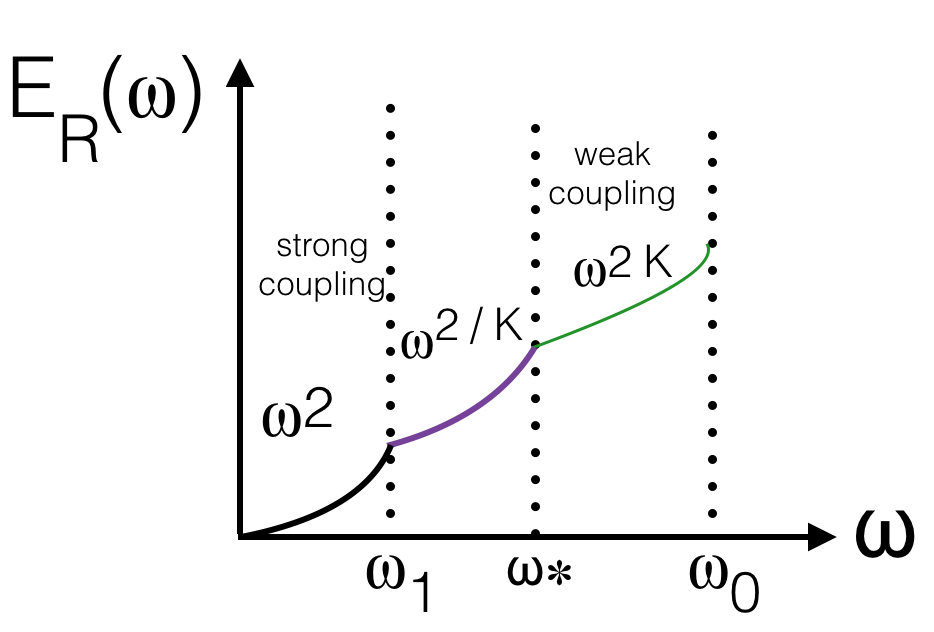}
\end{center}
\caption{\label{fig:fig3} (color online) Energy deposition as a function of frequency for a weak bare potential (see text). Depending on the frequency of the modulation
three absorption regime exist. Above the crossover frequency $\omega=\omega_\ast$ (dashed line), the backscattering term dominates with a powerlaw absorption $\omega^{2K}$.
Between $\omega_\ast$ and $\omega_1$  the backward scattering term dominates with an exponent  $2/K$ leading to $\omega^{2/K}$ behavior and below $\omega_1$ absorption is dominated by the forward scattering leading to an $\omega^2$ behavior.}
\end{figure}

The behavior of the energy deposition is depicted in Fig.~\ref{fig:fig3} for $K<1$.

\subsection{Strong bare potential} \label{sec:strong_coupling}

If the impurity potential is very large compared to the kinetic energy or interacting, one needs to start with the Hamiltonian (\ref{eq:barrier})
which corresponds to two semi-infinite chains coupled by a hopping term.

The energy modulation is thus given by
\begin{equation}\label{eq:eq18g}
\begin{split}
E_R&=\frac{\delta t_3^2 \pi}{4 u^4}\omega^4+\frac{1}{2}(\delta t_4)^2 \cos(\pi/K)\sin(\pi/K)\\& \Gamma(1-2/K) (u/\alpha)^{-2/K}\omega^{2/K}
\end{split}
\end{equation}
Where $\delta t_3=-\frac{2 a \sqrt{K}\delta V J^2}{\pi V_0^2}$, and $\delta t_4=\frac{2  J^2 \delta V\rho_0}{V_0^2}$, $a$ is the lattice spacing and $J$ the tunnelling.

The energy deposition rate shows again a power law behavior. The exponent is the same than the one corresponding to the renormalized backward scattering, discussed in the
previous section. Note however that the part corresponding to the forward scattering is now different. In this case the forward scattering contribute to an absorption going as
$\omega^4$ instead of $\omega^2$ for a weak potential and thus the energy absorption is dominated by the \emph{backscattering} potential,
as long as $0.5 < K < 1$. When $K < 0.5$ the system is again dominated by the absorption coming from the forwards scattering.

In the case $K > 1$ the tunnelling between the two half system will scale up as described the flow (\ref{eq: strong to weak1}). There is thus again a scale $\omega_\ast$ at which the tunnelling becomes of order one and then the Hamiltonian switches back to the one with a weak backscattering term. In that case the exponent becomes similar to the one obtained in Sec. \ref{sec:weakcoupling}
and is $2K$. So for $1 < K < 2$ the backscattering absorption dominates as very small frequencies, while the forward scattering dominates for $K > 2$.

Note that since the pre factors in the strong coupling limit for both backward and forward scattering are proportional to $1/V_0^2$, the energy deposition will be very small in comparison to the weak coupling case.

\section{Results and discussion} \label{sec:discussion}

Our results show that the time dependent modulation of a single impurity in an interacting bath of bosons or fermions allows to probe the TLL nature of the bath, the frequency of the modulation
and the interactions between particles serving as control parameters. Each regime is essentially characterized by a powerlaw behavior of the absorbed energy. As detailed in the previous sections
the precise power depends on: i) the forward and the backward scattering parts of the impurity potential; ii) the initial strength of the impurity potential compared to the energy scales of the bath
Hamiltonian; iii) the value of the interactions between the particles in the bath.
The results are summarized in Fig.~\ref{fig:fig2}.

Experiments in ultracold gases would thus be prime candidates to test the effects predicted there. Indeed several experiments with bulk modulation of the amplitude of the optical lattices
have already shown \cite{stoferle_tonks_optical, schori_absorption_optical, Griefdoublon_produnction_modulation} the interest of this technique. We here propose to do this modulation locally.
Since the modulation is local this has the advantage to be less sensitive to perturbations such as a confining harmonic potential. On the other hand the effect of the perturbation is now
much smaller of the order of $1/L$ where $L$ is the size of the system, and thus more difficult to measure.

Let us discuss separately the two cases of bosons and fermions.

\subsection{Bosonic systems}

Realizing interacting one dimensional bosonic systems in which modulation is possible has been already demonstrated \cite{stoferle_tonks_optical,Bloch_Higgs_mode_cold_atoms}.
Putting a local potential is possible either with a light blade \cite{Minardi_Loading_of_an_impurity_in_cold_atoms_by_light_blade} or in boson microscopes by locally addressing a single site \cite{Bloch_Boson_Microscope}. Measurement of the energy deposited in the system can be obtained either  by measuring
the momentum distribution \cite{stoferle_tonks_optical}  or in a microscope by looking at the number fluctuations \cite{Bloch_Boson_Microscope}. Especially in the last case, realizing
box potentials should also be possible.

For bosons with local interactions we have $K > 1$ and thus are on the right part of the Fig.~\ref{fig:fig2}. A weak potential should thus show an absorption dominated by the forward scattering potential, scaling essentially as $\omega^2$. It could be interesting if feasible to engineer a local potential so that the forward part of the potential would be much smaller than the backward
one. For example for a system with one boson every two site it would require a potential of the form $V_0$ on $x=-1$ and $-V_0$ on $x=0$. This would reinforce strongly the backscattering part and thus show the $\omega^{2K}$ absorption that would correspond to this component of the potential.

Alternatively one could get rid of the forward scattering component by going to a very strong on-site potential as detailed in Sec. \ref{strong_coupling}.
In such a case one would start with absorption scaling as $\omega^{2/K}$ which for weakly repulsive bosons for which $K$ can be large would correspond to a nearly constant absorption rate.
At the crossover $\omega_\ast$ the absorption would drop brutally since it would behave as $\omega^{2K}$. The crossover scale $\omega_\ast$ corresponding to the renormalization of the potential
in the \emph{static} part, should thus be visible with this method.

\subsection{Fermionic systems}

Shaking in fermionic systems has already been used for Hubbard-like Hamiltonians \cite{Griefdoublon_produnction_modulation} .
Instead of measuring the deposited energy, the growth of double occupancy allows for a more sensitive detection technique \cite{Griefdoublon_produnction_modulation, Tokuno_Doublon_production_rate_in_modulated_optical_lattices,Kollath_Phase_Modulation}. Fermionic systems allow also potentially to reach the regime $K < 1$.
However single component fermionic systems would require to have also finite range interactions which is for the moment still not too common in experiments. Reaching such a regime with long
range interactions can of course be also done with bosonic systems.
For fermions it is however simpler to use two component systems. In particular recent experiments have demonstrated the possibility to have fermionic microscopes
\cite{Zwierlein_Fermion_Microscope,Kuhr_Fermion_Microscope,ImmanuelBloch_Fermion_Microscope}
in which such two component fermionic systems can be manipulated. These experiments would be ideal testbeds for the effects described in the previous sections.

One could then explore potentially the same effects that the ones for bosons, namely the scaling $\omega^{2K}$ and $\omega^{2/K}$ of the absorbed energy.
In particular for a very large repulsion $U$ and a weak bare local potential one would expect an absorption $E_R \sim \omega^{3/2}$ since the limit
of the exponent $K_\rho$ for the Hubbard model is $K_\rho = 1/2$.

\section{conclusion} \label{sec:conclusion}

In this work, we have studied the energy deposition due to the periodic modulation in time of a local potential in an interacting one dimensional systems of bosons or fermions.
Using linear response and bosonization we have computed the  absorbed energy both for the regimes of a weak and a strong local potential. We find an absorption varying as a
powerlaw of the modulating frequency, with exponents that depend both on the interactions in the system and the nature and strength of the local potential.
The results are summarized in Fig.~\ref{fig:fig2}. This behavior allows to probe the TLL nature of the one dimensional interacting systems and we have discussed the experimental consequences
of potential experiments in ultracold gases.

Quite generally this study shows that the possibility to engineer local time dependent potentials allows to use the frequency of the modulation probe as a scale to explore the various renormalization regimes that the local potential would lead to.
The advantage of the technique is the fact that the modulation frequency can be precisely controlled in a large range of scales. The drawback is clearly that the measurement of the energy deposited is an effect varying as the inverse size of the system and thus difficult to measure. One could thus replace such a measurement by measurements of the \emph{local} density, which could potentially be achieved in systems such as the microscopes. Studies along that line are under way.

\begin{acknowledgments}
We would like to thank I. Bloch for interesting discussions.
This work is supported by the Swiss National Science Foundation under Division II.
\end{acknowledgments}

\appendix
\section{Linear response} \label{sec:linrep}

We give here a brief reminder of the energy deposited in a system by a periodic modulation of the form (\ref{eq:general})
The rate of energy deposition in a system
of Hamiltonian  $H=H_0+ a_0 \cos(\omega t) O$, is given by
\begin{equation}\label{eq:eq1A}
\begin{split}
\dot{E}&=\frac{\partial \bar{H}}{\partial t} \\
&=-\omega a_0\mathrm{sin}(\omega t)\overline{O(x=0,t)}
\end{split}
\end{equation}
  Where $H_0$ is a static part of Hamiltonian, $a_0$ is the amplitude of the time  dependent oscillation $\cos(\omega t)$ and $O$ is an operator coupled to the time dependent oscillation $\cos(\omega t)$,  within LRT, $ \bar{O}$ is given by
 \begin{equation}\label{eq:eq2A}
\begin{split}
\overline{O(t)}&=\langle O(t)\rangle_{H_0}-i a_0\int_{0}^{\infty} dt_1  \cos(\omega(t-t_1)) \\& \langle[O(t_1), O(0)]\rangle_{H_0}
\end{split}
\end{equation}
By using (\ref{eq:eq2A}), $\overline{O}$ is given by
\begin{equation}\label{eq:eq3A}
\begin{split}
\overline{O(t)}&=\langle O(t)\rangle_{H_0}-i a_0\int_{0}^{\infty} dt_1  \cos(\omega(t-t_1)) \\& \langle[O(t_1), O(0)]\rangle_{H_0}\\
\overline{O(t)}&=\langle O(t)\rangle_{H_0}+a_0\int_{0}^{\infty} dt_1  \cos(\omega(t-t_1))  X^R(t_1)\\
\overline{O(t)}&=\langle O(t)\rangle_{H_0}+\frac{a_0}{2}[e^{i\omega t}\chi^R(-\omega)+e^{-i\omega t}\chi^R(\omega)]\\
\end{split}
\end{equation}
where $\chi^R(t)=-i Y(t)\langle[O(t), O(0)]\rangle_{H_0} $ and $Y(t)$ is the Heaviside function.

By using (\ref{eq:eq3A}), (\ref{eq:eq1A}) can be expressed as
\begin{equation}\label{eq:eq4A}
\begin{split}
\dot{E}(t)&=-\omega a_0\mathrm{sin}(\omega t)\langle O(t)\rangle_{H_0}\\&-\frac{1}{2}a_0^{2} \omega\sin(\omega t)[ e^{i\omega t} \chi^R(-\omega)+e^{-i\omega t}  \chi^R(\omega)]
\end{split}
\end{equation}
We now calculate the cycle average of $\dot{E}(t)$ over the period of $T=2\pi/\omega$.
\begin{equation}\label{eq:eq5A}
\begin{split}
E_R&=\frac{1}{T} \int_0^T dt\dot{E}(t) \\
&=-\frac{1}{4 i}a_0^{2} \omega[-\chi^R_2(-\omega)+\chi^R(\omega)]\\
&=-\frac{1}{2}a_0^{2} \omega  \mathrm{Im}[\chi^R(\omega)]\\
\end{split}
\end{equation}

\section{Effective Hamiltonian for large V}\label{sec:effective hamiltonian}
In this section, we  derive the effective Hamiltonian for a system of hard core bosons  for a large V, in power of $1/V$. We consider, the site at $x=0$ is connected to sites at $x=-1$ and $x=1$ via tunnelling $J$, and nearest neighbour interaction $V_n$.
\begin{equation}\label {eq:eqE0}
H=H_1+H_2+H_3
\end{equation}
Where, $H_1$ and $H_2$  represent  the left and right parts of the Hamiltonian. $H_3$ is given by
\begin{equation}\label {eq:eqE0}
H_3=-J [\psi_{-1}^\dagger \psi_0+ \psi_{1}^\dagger \psi_0+ h.c] + V \psi_0^\dagger \psi_0+V_nn_0(n_{-1}+n_1)
\end{equation}
$H_3$ in the basis of $ |0,0,0\rangle,   |0,1,0\rangle,  |0,0,1\rangle,  |1,0,0\rangle,  |1,1,0\rangle,  $ $|1,0,1\rangle    |0,1,1\rangle ,  |1,1,1\rangle$
\begin{equation}\label {eq:eqE1} H_3=
\begin{bmatrix}
0 & 0 & 0 & 0 & 0 & 0 & 0 & 0 \\
0 & V & -J & -J & 0 & 0 & 0 & 0 \\
0 & -J & 0 & 0 & 0 & 0 & 0 & 0 \\
0 & -J & 0 & 0 & 0 & 0 & 0 & 0 \\
0 & 0 & 0 & 0 & V+V_n & -J & 0 & 0 \\
0 & 0 & 0 & 0 & -J & 0 & -J & 0 \\
0 & 0 & 0 & 0 & 0 & -J & V+V_n & 0 \\
0 & 0 & 0 & 0 & 0 & 0 & 0 & V+2V_n \\
\end{bmatrix}
\end{equation}
$H_3$ is composed of two block diagonal matrices depending on the total number of particles. We examine the part with one particle and two particles.
\begin{equation}\label {eq:eqE2} \tilde{H} =
\begin{bmatrix}
 V & -J & -J & 0 & 0 & 0  \\
 -J & 0 & 0 & 0 & 0 & 0  \\
 -J & 0 & 0 & 0 & 0 & 0  \\
 0 & 0 & 0 & V+V_n & -J & 0  \\
 0 & 0 & 0 & -J & 0 & -J  \\
 0 & 0 & 0 & 0 & -J & V+V_n \\
\end{bmatrix}
\end{equation}
$ \tilde{H}=\tilde{H}_t+\tilde{H}_V$, where $\tilde{H}_t$ contains only tunnelling, and $\tilde{H}_V=\tilde{H}-\tilde{H}_t$. In order to find the low energy effective Hamiltonian, we make a canonical transformation of $\tilde{H}$.
\begin{eqnarray*}
\overline{H} = W\tilde{H}W^{\dagger}&\simeq&\left(1+i\tilde{S}+\frac{(i)^2}{2}\tilde{S}^2\right)\tilde{H}\left(1-i\tilde{S}+\frac{(-i)^2}{2}\tilde{S}^2\right)\\
&\simeq&\tilde{H}+i[\tilde{S},\tilde{H}]+\frac{i^2}{2}[\tilde{S},[\tilde{S},\tilde{H}]]+\mathcal{O}(\tilde{S}^3)\\
\end{eqnarray*}
where $W=e^{i \tilde{S}}$. The canonical transformation does not change the energy but rotate the state $|\psi\rangle$ to $e^{i \tilde{S}} |\psi\rangle$ and we chose $\tilde{S}$  such that $\tilde{H}_t+i[\tilde{S},\tilde{H}_V]=0$
\begin{equation}\label {eq:eqE2} \tilde{S} =
\begin{bmatrix}
 0 & \frac{iJ}{V} &  \frac{iJ}{V} & 0 & 0 & 0 \\
 - \frac{iJ}{V} & 0 & 0 & 0 & 0 & 0 \\
 - \frac{iJ}{V} & 0 & 0 & 0 & 0 &0  \\
  0 & 0 & 0 &0 & \frac{iJ}{V_n+V} & 0\\
 0 & 0 & 0 & \frac{-iJ}{V_n+V} & 0 & \frac{-iJ}{V_n+V}\\
  0 & 0 & 0 & 0  &\frac{iJ}{V_n+V} & 0\\
\end{bmatrix}
\end{equation}
\begin{equation}
\overline{H}=\tilde{H_1}\oplus\tilde{H_2}
\end{equation}

\begin{equation}\label {eq:eqEffective} \overline{H} =
\begin{bmatrix}
 \frac{2 J^2}{V}+V & \frac{4J^3}{V^2} & \frac{4J^3}{V^2} &0&0&0  \\
 \frac{4J^3}{V^2} &  \frac{-J^2}{V} &  \frac{-J^2}{V}&0&0&0 \\
 \frac{4J^3}{V^2} & \frac{-J^2}{V} & \frac{-J^2}{V} &0&0&0 \\
 0&0&0&d+\frac{J^2}{V_n+V}&\frac{4J^3}{{(V_n+V)}^2}&\frac{J^2}{V_n+V}\\
 0&0&0&\frac{4J^3}{{(V_n+V)}^2}&\frac{-2J^2}{V_n+V}&\frac{4J^3}{({V_n+V)}^2}\\
 0&0&0&\frac{J^2}{V_n+V}&\frac{4J^3}{{(V_n+V)}^2}&d+\frac{J^2}{V_n+V}\\
\end{bmatrix}
\end{equation}
Where $d=V_n+V$.
\begin{equation}\label {eq:eqEffective} \tilde{H_1} =
\begin{bmatrix}
 \frac{2 J^2}{V}+V & \frac{4J^3}{V^2} & \frac{4J^3}{V^2}  \\
 \frac{4J^3}{V^2} &  \frac{-J^2}{V} &  \frac{-J^2}{V} \\
 \frac{4J^3}{V^2} & \frac{-J^2}{V} & \frac{-J^2}{V}  \\
\end{bmatrix}
\end{equation}
\begin{equation}\label {eq:eqEffective_1} \tilde{H_2} =
\begin{bmatrix}
 V_n+V+\frac{J^2}{V_n+V}&\frac{4J^3}{{(V_n+V)}^2}&\frac{J^2}{V_n+V}\\
 \frac{4J^3}{{(V_n+V)}^2}&\frac{-2J^2}{V_n+V}&\frac{4J^3}{({V_n+V)}^2}\\
 \frac{J^2}{V_n+V}&\frac{4J^3}{{(V_n+V)}^2}&V_n+V+\frac{J^2}{V_n+V}\\
\end{bmatrix}
\end{equation}
Since $V$ is very large, hence $1/V^2\simeq0$
\begin{equation} \label {eq:eqE3} \tilde{H_1} \simeq
\begin{bmatrix}
 \frac{2 J^2}{V}+V & 0 & 0  \\
 0 &  \frac{-J^2}{V} &  \frac{-J^2}{V} \\
 0 & \frac{-J^2}{V} & \frac{-J^2}{V}  \\
\end{bmatrix}
\end{equation}
\begin{equation}\label {eq:eqEffective_3} \tilde{H_2} =
\begin{bmatrix}
 V_n+V+\frac{J^2}{V_n+V}&0&\frac{J^2}{V_n+V}\\
 0&\frac{-2J^2}{V_n+V}&0\\
 \frac{J^2}{V_n+V}&0&V_n+V+\frac{J^2}{V_n+V}\\
\end{bmatrix}
\end{equation}
In the original Hamiltonian (\ref{eq:eqE1}), the $n_0=0$ sector  is coupled to $n_0=1$ sector via $J$ but after the canonical transformation the $n_0=0$ sector is completely decoupled from
the $n_0=1$  sector. The sector $n_0=0$ represents the low energy effective Hamiltonian. It is thus given by
\begin{equation}\label {eq:eqE4}
\begin{split}
 \overline{H} &=-\frac{J^2}{V}[|100\rangle \langle 100| +|001\rangle \langle 001|+|100\rangle \langle 001|+|001\rangle \langle 100|-\\& \frac{2J^2}{V+V_n}|101\rangle \langle 101|]
 \end{split}
\end{equation}
The full low energy Hamiltonian is thus
\begin{equation}\label {eq:eqE5}
\begin{split}
 H& =H_1+H_2-\frac{ J^2}{V}[|100\rangle \langle 100| +|001\rangle \langle 001|+\\& |100\rangle \langle 001|+|001\rangle \langle 100|- \frac{2J^2}{V+V_n}|101\rangle \langle 101|]
 \end{split}
\end{equation}
In second quantization (\ref{eq:eqE5}) can be written as
\begin{equation}\label {eq:eqE51}
\begin{split}
 H& =H_1+H_2+b(\psi_1(-a)^\dagger\psi_2(a)+h.c)-c(n_1(-a)+n_2(a))+\\& \tilde{V}n_1(-a)n_2(a)
 \end{split}
\end{equation}
Using (\ref{eq:eqE5})  and (\ref{eq:eqE51}) we find $b=-\frac{J^2}{V}$, $c=\frac{J^2}{V}$, $\tilde{V_n}-2c=- \frac{2J^2}{V+V_n}, \tilde{V_n}\simeq0$.

The bosonized form of $H$ is given by
\begin{equation}\label {eq:eqE6}
\begin{split}
H &= \frac{1}{2\pi}\int_{-\infty} ^{0} dx\Big[u K(\partial_x \theta_1(x))^2+\frac{u}{K} (\partial_x \phi_1(x))^2\Big] + \\& \frac{1}{2\pi}\int_{0} ^{\infty} dx\Big[ u K(\partial_x \theta_2(x))^2+\frac{u}{K} (\partial_x \phi_2(x))^2\Big]+\\&\frac {J^2}{\pi V} \partial_x \phi_1(-a)+\frac {J^2}{\pi V} \partial_x \phi_2(a)\\& -\frac {2J^2\rho_0}{ V}(\cos(2\phi_1(-a))+\cos(2\phi_2(a)))-\\& \frac{2J^2\rho_0}{V}\cos( \theta_1(-a)-\theta_2(a))
\end{split}
\end{equation}

At $x=0$, the number of particles is zero, which is imposed by $\phi_1(0)=\phi_2(0)=0$.
Now let us define $H$ as
\begin{equation}\label {eq:eqE7}
\begin{split}
H &= H_{10}+H_{20}+\frac {J^2}{\pi V} \partial_x \phi_1(-a)+\frac {J^2}{\pi V} \partial_x \phi_2(a)\\& -\frac {2J^2\rho_0}{ V}(\cos(2\phi_1(-a))-\cos(2\phi_2(a)))- \\& \frac{2 J^2\rho_0}{V}\cos( \theta_1(-a)-\theta_2(a))
\end{split}
\end{equation}
Where $H_{10}=\frac{1}{2\pi}\int_{-\infty} ^{0} dx\Big[u K(\partial_x \theta_1(x))^2+\frac{u}{K} (\partial_x \phi_1(x))^2\Big]$ and  $H_{20}=\frac{1}{2\pi}\int_{0} ^{\infty} dx\Big[u K(\partial_x \theta_2(x))^2+\frac{u}{K} (\partial_x \phi_2(x))^2\Big]$.

We define the fields $\theta=\frac{\tilde{\theta}}{\sqrt K}$ and $\phi=\sqrt K \tilde{\phi}$. Using these field $H_{10}$ and $H_{20}$ can be written as
\begin{equation}
 H_{10}=\frac{1}{2\pi}\int_{-\infty} ^{0} dx\Big[u (\partial_x \tilde{\theta}_1(x))^2+u (\partial_x \tilde{\phi}_1(x))^2\Big]
\end{equation}
and
\begin{equation}
 H_{20}=\frac{1}{2\pi}\int_{0} ^{\infty} dx\Big[u (\partial_x \tilde{\theta}_2(x))^2+u (\partial_x \tilde{\phi}_2(x))^2\Big]
\end{equation}
If we define $\tilde{\phi}, \tilde{\theta}$ in term of the chiral fields $\tilde{\phi}_L=\tilde{\phi}+\tilde{\theta}, \tilde{\phi}_R= \tilde{\theta}-\tilde{\phi}$,  the constraint $\phi_1(0)=\phi_2(0)=0$, imply that $\tilde{\phi}_L(x)=\tilde{\phi}_R(-x)$ \cite{giamarchi_book_1d}.
$H_{10}$ and $H_{20}$ are given in term of the new fields  $\tilde{\phi}_L$ and $\tilde{\phi}_R$ by
\begin{equation}\label {eq:eqE8}
\begin{split}
H_{10}=\frac{u}{4 \pi} \int_{-\infty}^{\infty} dx (\partial_x \tilde{\phi}_{R1}(x))^2\\
H_{20}=\frac{u}{4 \pi} \int_{-\infty}^{\infty} dx (\partial_x \tilde{\phi}_{L2}(x))^2
\end{split}
\end{equation}

In terms of $\tilde{\phi}_{R1}$ and $\tilde{\phi}_{L2}$ the term
$\partial_x \phi_1(-a)+ \partial_x \phi_2(a)$ becomes
\begin{equation}\label {eq:eqE9}
\begin{split}
\partial_x (\phi_1(-a)+  \phi_2(a))=\sqrt{K }\partial_x[ \tilde{\phi}_1(-a)+ \tilde{\phi}_2(a)]\\
=\frac{\sqrt{K }}{2}\partial_x[ \tilde{\phi}_{L1}(-a)-\tilde{\phi}_{R1}(-a)+\tilde{\phi}_{L2}(a)-\tilde{\phi}_{R2}(a) ]\\
=\frac{\sqrt{K }}{2}\partial_x[ \tilde{\phi}_{R1}(a)-\tilde{\phi}_{R1}(-a)+\tilde{\phi}_{L2}(a)-\tilde{\phi}_{L2}(-a) ]\\
\simeq \frac{\sqrt{K }}{2} 2 a \partial_x[  \partial_x\tilde{\phi}_{R1}(0)+ \partial_x\tilde{\phi}_{L2}(0) ]\\
=a \sqrt{K} \partial_x[  \partial_x\tilde{\phi}_{R1}(0)+ \partial_x\tilde{\phi}_{L2}(0) ]\\
\end{split}
\end{equation}

In the same way
\begin{equation}\label {eq:eqE10}
\begin{split}
\cos( \theta_1(-a)-\theta_2(a))=\cos\Big( \frac{1}{\sqrt K}(\tilde{\theta}_1(-a)-\tilde{\theta}_2(a))\Big)\\
=\cos\Big( \frac{1}{2 \sqrt K}(\tilde{\phi}_{L1}(-a)+\tilde{\phi}_{R1}(-a)-\tilde{\phi}_{L2}(a)-\tilde{\phi}_{R2}(a))\Big)\\
\simeq \cos\Big( \frac{1}{\sqrt K}(\tilde{\phi}_{R1}(0)-\tilde{\phi}_{L2}(0))\Big)
\end{split}
\end{equation}
and
\begin{equation}\label {eq:eqE11}
\begin{split}
\cos(\phi_1(-a))=\cos\Big(\frac{\sqrt{K}}{2}(\tilde{\phi}_{R1}(a)-\tilde{\phi}_{R1}(-a))\Big)\\
\cos(\phi_2(a))=\cos\Big(\frac{\sqrt{K}}{2}(\tilde{\phi}_{L2}(a)-\tilde{\phi}_{L2}(-a))\Big)
\end{split}
\end{equation}

Finally we define $\tilde{\tilde{\theta}}=\frac{\tilde{\phi}_{R1}+\tilde{\phi}_{L2}}{2}$ and $\tilde{\tilde{\phi}}=\frac{\tilde{\phi}_{L2}-\tilde{\phi}_{R1}}{2}$, and by using (\ref{eq:eqE8}),(\ref{eq:eqE9}), (\ref{eq:eqE10}), (\ref{eq:eqE11}), the Hamiltonian (\ref{eq:eqE7}) can be written as
\begin{equation}\label {eq:eqE12}
\begin{split}
H &= \frac{u}{2\pi}\int_{-\infty} ^{\infty} dx\Big[ (\partial_x \tilde{\tilde{\theta}}(x))^2+ (\partial_x \tilde{\tilde{\phi}}(x))^2\Big] +\\& \frac {2 a \sqrt{K}J^2}{\pi V} \partial_{x}^2 \tilde{\tilde{\theta}}(0) -\frac {2J^2\rho_0}{ V}\cos\Big(\frac{\sqrt K}{2}( \tilde{\tilde{\theta}}(a)-\tilde{\tilde{\theta}}(-a))\Big)\\& \cos\Big(\frac{\sqrt K}{2}( \tilde{\tilde{\phi}}(a)-\tilde{\tilde{\phi}}(-a))\Big)-  \frac{2 J^2\rho_0}{V}\cos\Big(\frac{ 2}{\sqrt K}( \tilde{\tilde{\phi}}(0))\Big)
\end{split}
\end{equation}
Since $V=V_0+\delta V \cos(\omega t)$, and $V_0 \gg \delta V$, this means that $1/V=1/V_0(1-\delta V/V_0 \cos(\omega t))=1/V_0-\delta V/V_0^2 \cos(\omega t)$.

\section{Calculation of correlation functions} \label{sec:correl}

In this section, we compute  response functions that are useful to obtain the energy deposited in the system.
Two types of correlation functions are required and denoted by $\chi_1^R(\omega)$ and $\chi_2^R(\omega)$ .
These correlation functions are defined below.
\begin{equation}\label{eq:eq1B}
\begin{split}
\chi_1(\tau)&=-\langle T_\tau\cos(2\phi(\tau))\cos(2\phi(0))\rangle_{H_0} \\
&=-\langle T_\tau\sin(2\phi(\tau))\sin(2\phi(0))\rangle_{H_0} \\
&=-\frac{1}{2} \langle T_\tau e^{2 i\phi(\tau)}e^{-2 i\phi(0)}\rangle_{H_0} \\
&=-\frac{1}{2} e^{-2K\log(u|\tau|/\alpha)} \\
&=-\frac{1}{2}(u|\tau|/\alpha)^{-2K}
\end{split}
\end{equation}
Where $T_\tau$ is time ordering operator and $\tau=it$ is imaginary time, t is real time.\\
Correlation function $\chi_1(\tau)$ in real time is given by
\begin{equation}\label{eq:eq2B}
\begin{split}
\chi_1^{T}(t)&=-\frac{1}{2}(u i t/\alpha)^{-2K} \\
&=-\frac{1}{2}(i)^{-2K}(u  t/\alpha)^{-2K} \\
&=-\frac{1}{2}(u  t/\alpha)^{-2K}(\cos(\pi K)-i\sin(\pi K))
\end{split}
\end{equation}
The retarded correlation function in real time is defined as
\begin{equation}\label{eq:eq3B}
\begin{split}
\chi_1^R(t)&=-i Y(t)\langle[\cos(2\phi(t)), \cos(2\phi(0))]\rangle_{H_0} \\
&=-Y(t)[2 \Im\chi_1^{T}(t)] \\
&=-Y(t)\sin(\pi K)(u t/\alpha)^{-2K}
\end{split}
\end{equation}

\subsection{Fourier transformation of $\chi^R(t)$}

Fourier transformation of $\chi_1^R(t)$ is expressed as
\begin{equation}\label{eq:eq4B}
\begin{split}
\chi_1^R(\omega)&=\int_{-\infty}^{\infty} dt\ e^{i \omega t} \chi^R(t) \\
&=-\int_{0}^{\infty} dt\ e^{i \omega t}\sin(\pi K)(u t/\alpha)^{-2K}
\end{split}
\end{equation}
The above integral can be evaluated as
\begin{eqnarray}
\int_{0}^{\infty}dt\ e^{i \omega t}\sinh(\pi t/\beta)^{-2K}=\nonumber \\2^{2K}\frac{\beta}{2\pi}  B\Big(\frac{-i\beta\omega}{2\pi}+K,1-2K)
\label{eq:eq5B}
\end{eqnarray}
As $\beta\rightarrow\infty$, $\sinh(\pi t/\beta)=\pi t/\beta\Rightarrow \beta/\pi\sinh(\pi t/\beta)= t$
\begin{eqnarray}
\int_{0}^{\infty}dt\ e^{i \omega t}(\beta/\pi)^{-2K}\sinh(\pi t/\beta)^{-2K}= \nonumber\\2^{2K-1}\Big(\frac{\beta}{\pi}\Big)^{-2K+1} B\Big(\frac{-i\beta\omega}{2\pi}+K,1-2K\Big)
\label{eq:eq6B}
\end{eqnarray}
Equation (\ref{eq:eq5B}) can be expressed as
\begin{equation}\label{eq:eq7B}
\begin{split}
\int_{0}^{\infty}dt\ e^{i \omega t}(t)^{-2K}&=2^{2K-1}\Big(\frac{\beta}{\pi}\Big)^{-2K+1} \\& \quad
B\Big(\frac{-i\beta\omega}{2\pi}+K,1-2K\Big) \\
&\simeq 2^{2K-1}\Big(\frac{\beta}{\pi}\Big)^{-2K+1}\Gamma(1-2K) \\ & \quad \Big(\frac{-i\beta \omega}{2 \pi}\Big)^{2K-1} \\
&=(-i\omega)^{2K-1}\Gamma(1-2K)\\
&=e^{-i\pi(K-1/2)}\Gamma(1-2K)\omega^{2K-1}
\end{split}
\end{equation}
Equation (\ref{eq:eq4B}) can be written as
\begin{equation}\label{eq:eq8B}
\begin{split}
\chi_1^R(\omega)&= -\sin(\pi K)(u/\alpha )^{-2K}e^{-i\pi(K-1/2)} \\& \Gamma(1-2K)\omega^{2K-1} \\
\Im[\chi_1^R(\omega)]&=\sin(\pi K)(u/\alpha )^{-2K}\sin(\pi(K-1/2)) \\& \Gamma(1-2K)\omega^{2K-1}
\end{split}
\end{equation}
Response function of $\partial_x\phi(0,0)$ is given by
\begin{equation}\label{eq:eq9B}
\begin{split}
\chi_2^R(t)&=-i Y(t)\langle[\partial_x\phi(0,t),\partial_x\phi(0,0) ]\rangle_{H_0}\\
\chi_2^R(\omega)&= -\frac{u K}{2}\int dk \frac{k^2}{-(\omega+i\delta)^2+(uk)^2} \\
\Im[\chi_2^R(\omega)]&=-\frac{u K \pi}{4 \omega}\int dk k^2\delta\Big(\frac{-\omega^2+(uk)^2}{2\omega}\Big) \\
\Im[\chi_2^R(\omega)]&=-\frac{K\pi\omega}{2u^2}
\end{split}
\end{equation}

\section{Calculation of correlation functions when $V_0$ is relevant$(K<1)$} \label{ap:instanton}

In this section, we will derive the two point correlation function $\langle \phi(q_1)  \phi(q_2)\rangle$ in term of a local field $\phi(0,\tau)$. We define the local field $\phi_0(\tau)$, acting at $x=0$ as
\begin{equation}\label {eq:eqD1}
\phi(0,\tau)=\phi_0(\tau)
\end{equation}
By using (\ref{eq:eqD1}), $R(q_1,q_2)=\langle \phi(q_1)  \phi(q_2)\rangle$ can be written as
\begin{equation}\label {eq:eqD2}
\begin{split}
 R(q_1,q_2)&=\frac{\int \mathcal{D}\phi  \mathcal{D}\phi_0  \mathcal{D}\lambda  \phi(q_1)  \phi(q_2) e^{(-S-i\int d\tau \lambda(\tau)[\phi_0(\tau)-\phi(0,\tau)])}}{\int \mathcal{D}\phi  \mathcal{D}\phi_0  \mathcal{D}\lambda  e^{(-S-i\int d\tau \lambda(\tau)[\phi_0(\tau)-\phi(0,\tau)])}}
\end{split}
\end{equation}
$S$ is action of the system, and $\lambda$ is a Lagrange multiplier. Now define $-S-i\int d\tau \lambda(\tau)[\phi_0(\tau)-\phi(0,\tau)]$ in terms of the frequency($\omega$) and momentum($k$).
\begin{equation}\label {eq:eqD3}
\begin{split}
I&=S+i\int d\tau \lambda(\tau)[\phi_0(\tau)-\phi(0,\tau)]\\
&=\frac{1}{2\pi K u}\frac{1}{\beta\Omega} \sum_q(\omega_n^2+u^2k^2)  \phi^\ast(q)\phi(q)+\\& i\Big[\frac{1}{\beta}\sum_{\omega_n}\lambda^\ast(\omega_n)\phi_0(\omega_n)- \frac{1}{\beta\Omega}\sum_{q}\lambda^\ast(\omega_n)\phi(q)\Big]+ \\& 2V_0\rho_0\int d\tau \cos(2\phi_0(\tau))\\
&=\frac{1}{2\pi K u}\frac{1}{\beta\Omega}  \sum_q(\omega_n^2+u^2k^2)  (\phi^\ast(q)-\frac{i\pi u K}{(\omega_n^2+u^2k^2)}\lambda^\ast(\omega_n)) \\& (\phi(q)-\frac{i\pi u K}{(\omega_n^2+u^2k^2)}\lambda(\omega_n))+ \frac{\pi K}{4\beta} \\& \sum_{\omega_n}\frac{1}{|\omega_n|}(\lambda^\ast(\omega_n)+\frac{2 i |\omega_n|}{\pi K} \phi^\ast_0(\omega_n))(\lambda(\omega_n)+\frac{2 i |\omega_n|}{\pi K} \phi_0(\omega_n))\\& +\frac{1}{\beta}\sum_{\omega_n}\frac{|\omega_n|}{\pi K}\phi^\ast_0(\omega_n)\phi_0(\omega_n)+ 2V_0\rho_0\int d\tau \cos(2\phi_0(\tau))
\end{split}
\end{equation}
By using (\ref{eq:eqD3}), (\ref{eq:eqD2}) can be written as
\begin{equation}\label {eq:eqD4}
\begin{split}
R(q_1,q_2)&=\frac{\pi K \beta\Omega u}{\omega_{n_1}^2+u^2k_1^2}\delta_{q_1,-q_2}-\\& \frac{\pi^2u^2K^2}{(\omega_{n_1}^2+u^2k_1^2)(\omega_{n_2}^2+u^2k_2^2)}\frac{2\beta|\omega_{n_1}|}{\pi K}\delta_{\omega_{n_1},-\omega_{n_2}}+\\& \frac{\pi^2u^2K^2}{(\omega_{n_1}^2+u^2k_1^2)(\omega_{n_2}^2+u^2k_2^2)}\frac{4 |\omega_{n_1}| |\omega_{n_2}|}{\pi^2K^2}\\& \frac{\int \mathcal{D}\phi_0 \phi_0(\omega_{n_1})\phi_0(\omega_{n_2})e^{-\frac{1}{\beta\pi K} \sum_{\omega_n}|\omega_n|\phi^\ast_0(\omega_{n})\phi_0(\omega_{n})-S_I}} {\int \mathcal{D}\phi_0 e^{-\frac{1}{\beta\pi K} \sum_{\omega_n}|\omega_n|\phi^\ast_0(\omega_{n})\phi_0(\omega_{n})- S_I}}
\end{split}
\end{equation}
Where $S_I=2V_0\rho_0\int d\tau \cos(2\phi_0(\tau))$
\begin{equation}\label {eq:eqD5}
\begin{split}
-\langle \nabla \phi(0,\tau)\nabla \phi(0,0)\rangle&=-{\Big(\frac{1}{\beta\Omega}\Big)}^2\sum_{q_1,q_2}(-k_1 k_2) e^{(-i \omega_{n_1}\tau)} \\& \langle\phi(q_1)\phi(q_2)\rangle
\end{split}
\end{equation}
By using  (\ref{eq:eqD3}),  (\ref{eq:eqD5}) can be written as
 \begin{equation}\label {eq:eqD6}
\begin{split}
-\langle \nabla \phi(0,\tau)\nabla \phi(0,0)\rangle&=-{\Big(\frac{1}{\beta\Omega}\Big)}^2\sum_{q_1,q_2}(-k_1 k_2) e^{(-i \omega_{n_1}\tau)} \\& \frac{\pi K \beta\Omega u}{\omega_{n_1}^2+u^2k_1^2}\delta_{q_1,-q_2}
\end{split}
\end{equation}

Fourier transformation of  (\ref{eq:eqD6}) can be written as
 \begin{equation}\label {eq:eqD7}
\begin{split}
-\langle \nabla \phi(0,\tau)\nabla \phi(0,0)\rangle(\omega_n)&=-{\Big(\frac{1}{ \Omega}\Big)}\sum_{k}(k^2)  \frac{\pi K  u}{\omega_{n}^2+u^2k^2}\\
-\Im(\langle \nabla \phi(0,\tau)\nabla \phi(0,0)\rangle(\omega))&=-\frac{K\pi\omega}{2 u^2}
\end{split}
\end{equation}

\subsection{Dilute instanton approximation}

By using (\ref{eq:eqD3}), the effective action in term of the local field $\phi_0$ is given by
\begin{equation}\label {eq:eqD8}
\begin{split}
S&=\frac{1}{\beta\pi K} \sum_{\omega_n}|\omega_n|\phi^\ast_0(\omega_{n})\phi_0(\omega_{n})+2V_0\rho_0\int d\tau \cos(2\phi_0(\tau))
\end{split}
\end{equation}
The action diverges for the large frequency, and to overcome this problem, we add a mass term $\frac{1}{2}M{(\partial_\tau\phi)}^2$\cite{giamarchi_book_1d,kane_luttinger_impurity} to the action.
\begin{equation}\label {eq:eqD9}
\begin{split}
S&=\frac{1}{\beta\pi K} \sum_{\omega_n}|\omega_n|\phi^\ast_0(\omega_{n})\phi_0(\omega_{n})+2V_0\rho_0\int d\tau \cos(2\phi_0(\tau))\\& +\int d\tau \frac{1}{2}M{(\partial_\tau\phi)}^2
\end{split}
\end{equation}
For a very large $V_0$, the partition function is dominated by  a trajectory, which minimizes the action $2V_0\rho_0\int d\tau \cos(2\phi_0(\tau))+ \int d\tau \frac{1}{2}M{(\partial_\tau\phi)}^2$, and solution is given by \cite{giamarchi_book_1d, furusaki_luttinger_impurity}
\begin{equation}\label {eq:eqD10a}
\begin{split}
\frac{M}{2}\Big(\frac{d\phi(\tau)}{d\tau}\Big)^2=\cos(2\phi(\tau))-1
\end{split}
\end{equation}
\begin{equation}\label {eq:eqD10}
\begin{split}
\tilde{\phi}(\tau)&=\pi/2+2\tan^{-1}[\tanh[\sqrt{2V_0\rho_0/M}\tau]]
\end{split}
\end{equation}
For $V_0>>0$, $\partial_\tau\tilde{\phi}(\tau)\simeq \delta(\tau)$, where $\delta(\tau)$ is delta function. The general solution of the field $\phi_0(\tau)$  is given by the linear combination of $\tilde{\phi}(\tau)$,
\begin{equation}\label {eq:eqD11}
\begin{split}
{\phi_0}(\tau)&=\sum_i\epsilon_i\tilde{\phi}(\tau-\tau_i)
\end{split}
\end{equation}
Where $\epsilon_i= \pm1$, and $\sum_i\epsilon_i=0$.
By using (\ref{eq:eqD11}) and (\ref{eq:eqD9}), partition function of the system  is given by \cite{giamarchi_book_1d}.\\
\begin{equation}\label {eq:eqD11a}
\begin{split}
Z&=\sum_{p=0}^{\infty} \Delta_{0}^{2p}\sum_{\epsilon_1=\pm 1...\epsilon_{2 p}=\pm 1}\int_{0}^{\infty}d\tau_{2p}\int_{0}^{\tau_{2p}}d\tau_{2p-1}....\int_{0}^{\tau_2}d\tau_{1}\\& e^{2/K\sum_{i>j}\epsilon_i\epsilon_j\log(|\tau_i-\tau_j|/\delta)}
\end{split}
\end{equation}
Where $ \Delta_{0}=e^{-4\sqrt{2\rho_0V_0M}}$, $\delta$ is short time cutoff. Since, in the strong coupling limit $V_0>>1$, so the  large number of instantons will give small contribution to the partition function because of the pre-factor $ \Delta_{0}=e^{-4\sqrt{2\rho_0V_0M}}$, and the dominant contribution is given by one instanton and one anti-instanton.\\
For our calculation, we  consider one instanton($\epsilon_1= 1$) and one anti-instanton($\epsilon_2= -1$). The field $\phi$ is given by
\begin{equation}\label {eq:eqD12}
\begin{split}
{\phi_0}(\tau)&=\tilde{\phi}(\tau-\tau_1)-\tilde{\phi}(\tau-\tau_2)
\end{split}
\end{equation}
$R_1(\tau)=\langle\cos({2\phi_0}(\tau))\cos({2\phi_0}(0))\rangle$
\begin{equation}\label {eq:eqD13}
\begin{split}
R_1(\tau)&=\Big(1+e^{-8\sqrt{2\rho_0V_0M}}\int_{-\beta/2}^{\beta/2} d\tau_1 \\& \int_{-\beta/2}^{\beta/2} d\tau_2  \cos(2\tilde{\phi}(\tau-\tau_1)-2\tilde{\phi}(\tau-\tau_2)) \\& \cos(2\tilde{\phi}(-\tau_1)-2\tilde{\phi}(-\tau_2)) {(|\tau_1-\tau_2|/\delta)}^{-2/K}\Big)\\&\Big(1+e^{-8\sqrt{2\rho_0V_0M}}\int_{-\beta/2}^{\beta/2} d\tau_1 \\& \int_{-\beta/2}^{\beta/2} d\tau_2 {(|\tau_1-\tau_2|/\delta)}^{-2/K}\Big)^{-1}\\
&\simeq - \frac{M^2}{2} e^{-8\sqrt{2\rho_0V_0M}} (|\tau|/\delta)^{-2/K}
\end{split}
\end{equation}
Where $\beta=1/T$, T is temperature.
By using appendix B, imaginary part of $\langle\cos({2\phi_0}(\tau))\cos({2\phi_0}(0))\rangle(\omega)$ is given by
\begin{equation}\label {eq:eqD14}
\begin{split}
\Im[R_1(\tau)](\omega)&= -M^2 e^{-8\sqrt{2\rho_0V_0M}}\cos(\pi/K)  \\& \sin{(\pi/K)} \Gamma(1-2/K)(\delta \ \omega)^{2/K-1}
\end{split}
\end{equation}

\bibliography{Time_Dependent}

\end{document}